\documentstyle[12pt]{article}

\begin{document}

\pagestyle{plain}

\begin{center}
{\bf TOTAL ABSORPTION IN FINITE\ TIME IN\ AN\ $i\delta $  POTENTIAL}

\medskip

by

\medskip

A. Marchewka\footnote{%
e-mail: ntmarche@weisemail.Weizmann.ac.il}\\[0pt]
Weizmann Institute of Science, Rehovot 76100, Israel\\[3mm]

Z. Schuss\footnote{%
e-mail: schuss@post.tau.ac.il}\\[0pt]
Department of Mathematics\\[0pt]

Tel-Aviv University, Tel-Aviv 69978, Israel\\[1cm]

{\bf ABSTRACT}
\end{center}

\vspace{3mm} \noindent We consider the evolution of Green's function of the
one-dimensional Schr\"odinger equation in the presence of the complex
potential $-ik\delta(x)$. Our result is the construction of an explicit
time-dependent solution which we use to calculate the time-dependent
survival probability of a quantum particle. The survival probability decays
to zero in finite time, which means that the complex delta potential well is
a total absorber for quantum particles. This potential can be interpreted as
a killing measure with infinite killing rate concentrated at the origin.

\newpage \pagestyle{plain}

The fundamental question of the existence of a totally absorbing potential
that absorbs a particle in finite time with probability 1 has been the
subject of extensive discussion in the literature \cite{Allcock1}, \cite
{Rubio}. It was claimed that no perfectly absorbing potential can exist. To
study this question, we consider the model experiment of releasing electrons
at an absorbing plane placed at the origin. This model has many uses, e.g.,
in scattering theory, solid state physics, surface dynamics, physical
chemistry and so on \cite{MugaCPL}-\cite{Patrici}. The problem can be
described \cite{Patrici} by the one-dimensional Schr\"odinger equation with
the complex potential $-ik\delta(x),\ (k>0)$. The description proposed in 
\cite{Patrici} considers a time-independent monochromatic wave at $-\infty$
that is measured at $+\infty$ (on the other side of the absorbing plane).
This is the standard time-independent description of scattering of plane
waves on an absorbing plane. The result of \cite{Patrici} is that up to $%
25\% $ of the incident current is absorbed by the plane.

A related time-independent description of this problem is due to Muga {\em %
et al.} \cite{MugaPRL}, \cite{MugaJPA}. In this description a complex
potential of finite, though arbitrarily small support, absorbs totally any
discrete set of plane waves. This precludes the possibility of absorbing a
normalizable wave packet.

In our time-dependent description of the same experiment electrons are
released at the absorbing plane one at a time. Their survival probability, $%
S(t)$, is defined as the relative number of electrons that have not been
absorbed by the surface by time $t$ since their release. We show that in
this description there is a critical time beyond which the survival
probability of an electron vanishes. This means that the given potential
leads to an eventual total absorption of all electrons released in finite
time with probability 1. This is in contrast to the assertion of the
time-independent model of this experiment. We also show that the probability
density on the surface converges in time to a finite limit, though the total
probability in space decays in time.

This description of an absorbing surface is different from the one proposed
in \cite{PLA} and \cite{Survival}. The description in \cite{PLA}, \cite
{Survival} assumes that the surface absorbs all Feynman trajectories at the
moment they reach the absorbing surface. The difference in the results of
the two time-dependent descriptions is that in the present description the
wave function can propagate across the surface while in the description
proposed in \cite{PLA} and \cite{Survival} it cannot. While the result of 
\cite{PLA}, \cite{Survival} is that the survival probability decays at and
exponential rate proportional to the absorption current at the absorbing
plane, the result of the present description is that the survival
probability decays in finite time.\newline

\noindent {\bf Free Brownian motion with $\delta$-killing}\newline

Consider the Brownian motion on the line with a killing measure $k\delta
\left( x\right) $ per unit time. The transition probability density
satisfies the diffusion equation 
\begin{eqnarray}
p_{t}=Dp_{xx}-\kappa \delta \left( x\right) p, \label{de}
\end{eqnarray}
or equivalently,
\[
p_{t}\left( x,t\right) =Dp_{xx}\left( x,t\right) -\kappa \delta \left(
x\right) p\left( 0,t\right) , 
\]
and the initial condition 
\begin{eqnarray}
p\left( x,0\right) =\delta \left( x-x_{0}\right) . \label{ic}
\end{eqnarray}
The survival probability is given by 
\[
S\left( t\right) =\int p\left( x,t\right) \,\,dx, 
\]
The diffusion equation gives 
\[
\frac{d}{dt}\int p\left( x,t\right) \,\,dx=-\kappa p\left( 0,t\right) , 
\]
hence 
\[
S\left( t\right) =1-\kappa \int_{0}^{t}p\left( 0,t\right) \,dt. 
\]
Changing variables to $Dt=\tau $, the solution of the initial value problem (%
\ref{de}), (\ref{ic}) satisfies 
\begin{eqnarray*}
p\left( x,\tau \right) &=&\int_{-\infty }^{\infty }\frac{1}{2\sqrt{\pi \tau }%
}\exp \left\{ -\frac{\left( x-y\right) ^{2}}{4\tau }\right\} \delta \left(
y-x_{0}\right) \,dy- \\
&& \\
&&\int_{0}^{\tau }\frac{1}{2\sqrt{\pi \left( \tau -\sigma \right) }}%
\int_{-\infty }^{\infty }\exp \left\{ -\frac{\left( x-y\right) ^{2}}{4\left(
\tau -\sigma \right) }\right\} \frac{\kappa }{D}\delta \left( y\right)
p\left( y,\sigma \right) \,dy\,d\sigma ,
\end{eqnarray*}
and in particular it satisfies the integral equation 
\begin{eqnarray}
p\left( 0,\tau \right) =\frac{1}{2\sqrt{\pi \tau }}\exp \left\{ -\frac{%
x_{0}^{2}}{4\tau }\right\} -\int_{0}^{\tau }\frac{1}{2\sqrt{\pi(\tau
-\sigma)}}\frac{\kappa}{D}p(0,\sigma)\,d\sigma . \label{ieq}
\end{eqnarray}
The Laplace transform of the integral equation (\ref{ieq}) is 
\[
\hat{p}(0,s)=\frac{e^{-x_{0}\sqrt{s}}}{2\sqrt{s}}-\frac{\kappa \hat{p}(0,s)}{%
2D\sqrt{s}}, 
\]
hence \cite[p.1026, eq.(29.3.88)]{Abramowitz}, 
\begin{eqnarray}
p(0,\tau ) &=&\frac{1}{2}{\cal L}^{-1}\left( \frac{e^{-x_{0}\sqrt{s}}}{\sqrt{%
s}+\frac{\kappa }{2D}}\right)  \nonumber \\
&&  \label{p0t} \\
&=&\frac{1}{2}\left\{ \frac{1}{\sqrt{\pi \tau }}\exp \left\{ -\frac{x_{0}^{2}%
}{4\tau }\right\} -\frac{\kappa }{2D}e^{\frac{\kappa }{2D}x_{0}}e^{\frac{%
\kappa ^{2}}{4D^{2}}\tau }\left[ 1-\mbox{erf}\left( \frac{\kappa }{2D}\sqrt{%
\tau }+\frac{x_{0}}{2\sqrt{\tau }}\right) \right] \right\}  \nonumber \\
&\sim &\frac{1}{2}\frac{\kappa x_{0}+2D}{\kappa ^{2}\tau ^{3/2}}\frac{D}{%
\sqrt{\pi }}\quad \mbox{for \quad }\tau \rightarrow \infty .
\end{eqnarray}
It follows that the large $\tau $ limit of $p\left( 0,\tau \right) $ is 
\[
\lim_{s\rightarrow 0}\frac{se^{-x_{0}\sqrt{s}}}{2\sqrt{s}+\frac{\kappa }{D}}%
=0. 
\]
The limit of the integral of $\frac{\kappa }{D}p\left( 0,\tau \right) $ is 
\[
\lim_{\tau \rightarrow \infty }\frac{\kappa }{D}\int_{0}^{\tau }p\left(
0,u\right) \,du=\lim_{s\rightarrow 0}\frac{\kappa }{D}\hat{p}%
(0,s)=\lim_{s\rightarrow 0}\frac{\kappa }{D}\frac{e^{-x_{0}\sqrt{s}}}{2\sqrt{%
s}+\frac{\kappa }{D}}=1. 
\]
It follows that the survival probability decays to zero.

The survival probability decays as 
\[
\frac{\kappa }{D}\int_{\tau }^{\infty }p\left( 0,\tau \right) \,d\tau =\frac{%
x_{0}+2D/\kappa }{\sqrt{\pi \tau }}+O\left( \frac{1}{\tau ^{3/2}}\right)
\quad \mbox{for\quad }\tau >>1. 
\]

\bigskip\ 

\noindent {\bf The survival probability in Schr\"odinger's equation with $%
-ik\delta$ potential}\newline

The wave function in the presence of the potential $-ik\delta (x)$ can be
expressed as the Feynman integral with {\em killing} rate $k\delta (x)$ \cite
{book}. This means that at each time step in the construction of the Feynman
integral the wave amplitude is discounted by the factor $1-\frac{1}{2}%
k\delta (x(t))\,\Delta t$. The $\delta $-function can be interpreted as the
usual limit of high and narrow rectangular potential whose area is 1. We
recall that the limit of a high rectangular potential with finite width
leads to total reflection rather than total absorption. This discount factor
corresponds to the additional factor 
\[
\exp \left\{ -\frac{k}{2}\delta (x_{j})\,\Delta t\right\} 
\]
in the propagator. The resulting Schr\"{o}dinger equation contains the
imaginary potential $-\displaystyle\frac{ik}{2}\delta (x)$, 
\begin{equation}
i\hbar \psi _{t}=-\frac{\hbar ^{2}}{2m}\Delta \psi +V(x)\psi -\frac{ik}{2}%
\delta (x)\psi .  \label{schreq}
\end{equation}
We assume for simplicity that 
\begin{equation}
\psi \left( y,0\right) =\delta \left( y-x_{0}\right)  \label{icS}
\end{equation}

When the initial condition is square-integrable, we obtain from (\ref{schreq}%
) that 
\[
\frac{d}{dt}\int |\psi (x,t)|^{2}\,dx=-\frac{k}{\hbar }\left| \psi \left(
0,t\right) \right| ^{2}, 
\]
so that the survival probability is given by 
\begin{equation}
S(t)=\int |\psi (x,t)|^{2}\,dx=1-k\int_{0}^{t}|\psi (0,t)|^{2}\,dt.
\label{Survival}
\end{equation}

To evaluate $S(t)$, we have to determine the decay rate of $\psi (0,t)$. To
do so, we begin with Green's function of the Schr\"{o}dinger equation 
\[
i\hbar G_{t}=-\frac{\hbar ^{2}}{2m}\Delta _{x}G+V(x)G 
\]
and the initial condition 
\[
\lim_{t\rightarrow 0}G(x,y,t)=\delta (x-y). 
\]
In the case of free propagation 
\begin{equation}
G(x,y,t)=\sqrt{\frac{m}{2\pi \hbar it}}\exp \left\{ -\frac{m(x-y)^{2}}{%
2\hbar it}\right\} .  \label{freeprop}
\end{equation}
\ \ \ \ \ \ \ \ \ \ 

The solution of (\ref{schreq}) can be written as 
\begin{eqnarray}
\psi (x,t) &=&\int G(x,y,t)\psi (y,0)\,dy-\frac{ik}{2}\int_{0}^{t}\int
G(x,y,t-s)\delta (y)\psi (y,s)\,dy\,ds  \nonumber \\
&&  \label{psixt} \\
&=&\int G(x,y,t)\psi (y,0)\,dy-\frac{ik}{2}\int_{0}^{t}G(x,0,t-s)\psi
(0,s)\,ds.  \nonumber
\end{eqnarray}
With the initial condition (\ref{icS}), we obtain 
\begin{equation}
\psi (0,t)=\int G(0,y,t)\psi (y,0)\,dy-\frac{ik}{2}\int_{0}^{t}G(0,0,t-s)%
\psi (0,s)\,ds.  \label{inteq}
\end{equation}

Equation (\ref{inteq}) is an integral equation for the function $\phi
(t)=\psi (0,t),$ that can be written as 
\begin{equation}
\phi (t)=f(t)-\int_{0}^{t}K(t-s)\phi (s)\,ds,  \label{phieq}
\end{equation}
where 
\begin{eqnarray}
f(t) &=&G(0,x_{0},t)=\sqrt{\frac{m}{2\pi \hbar it}}\exp \left\{ -\frac{%
mx_{0}^{2}}{2\hbar it}\right\}  \nonumber \\
&&  \label{ftKt} \\
K(t) &=&\frac{ik}{2}G(0,0,t)=\frac{k}{2}\sqrt{\frac{im}{2\pi \hbar t}}. 
\nonumber
\end{eqnarray}
The Laplace transform of the solution of eq.(\ref{ftKt}) is given by 
\[
\hat{\phi}(s)=\frac{\hat{f}(s)}{1+\hat{K}(s)}. 
\]
where 
\[
\hat{f}(s)=\sqrt{\frac{m}{2\hbar is}}\int e^{-\sqrt{\frac{2my^{2}}{\hbar i}s}%
}\psi (y,0)\,dy 
\]
and 
\[
\hat{K}(s)=\frac{k}{2}\sqrt{\frac{im}{2\hbar s}}. 
\]
The exact expression for $\phi \left( t\right) $ is obtained from eq.(\ref
{p0t}) with the values 
\[
\tau =Dt,\quad D=\frac{i\hbar }{2m},\quad \kappa =\frac{k}{2\hbar }. 
\]
It is given by 
\begin{eqnarray}
\phi \left( t\right) &=&\frac{\sqrt{m}}{\sqrt{2i\pi \hbar t}}\int_{-\infty
}^{\infty }e^{\frac{1}{2}i\frac{y^{2}}{\hbar t}m}\psi (y,0)\,dy+  \nonumber
\\
&&  \label{phiti} \\
&&\frac{1}{4}i\frac{k}{\hbar ^{2}}me^{-\frac{1}{8}i\frac{k^{2}}{\hbar ^{3}}%
mt}\int_{-\infty }^{\infty }e^{-\frac{1}{2}i\frac{k}{\hbar ^{2}}my}\times 
\nonumber \\
&&  \nonumber \\
&&\left[ 1-\mbox{erf}\left( -\frac{1}{4}i\frac{k}{\hbar ^{2}}m\sqrt{2}\sqrt{%
\left( i\hbar \frac{t}{m}\right) }+\frac{1}{2}y\frac{\sqrt{2}}{\sqrt{\left(
i\hbar \frac{t}{m}\right) }}\right) \right] \psi (y,0)\,dy.  \nonumber
\end{eqnarray}
Choosing 
\[
\psi (y,0)\,=\frac{1}{\sqrt{2\pi }a}\exp \left( \frac{-\left( y-x_{0}\right)
^{2}}{2a^{2}}\right) , 
\]
the first integral is 
\[
\left( 1-i\right) \frac{\sqrt{m}}{2\sqrt{\left( -ima^{2}+\hbar t\right) }%
\sqrt{\pi }}\exp \left( \frac{1}{2}ix_{0}^{2}\frac{m}{-ima^{2}+\hbar t}%
\right) . 
\]
To evaluate the second integral, we note that at the point $y=-\frac{1}{2}k%
\frac{t}{\hbar }\left( 1-\frac{1}{\sqrt{t}}\right) $ the argument of the
error function vanishes, so it cannot be expanded uniformly for large $t$.
Therefore, we break the line into the segments $-\infty <y<-\frac{1}{2}k%
\frac{t}{\hbar }\left( 1-\frac{1}{\sqrt{t}}\right) +\alpha \sqrt{t}$, and$%
\quad -\frac{1}{2}k\frac{t}{\hbar }\left( 1-\frac{1}{\sqrt{t}}\right)
+\alpha \sqrt{t}$ $<y<\infty $, where $\alpha $ is a positive constant to be
chosen. The integral over the first interval can be estimated by 
\begin{eqnarray*}
&&\int_{-\infty }^{-\frac{1}{2}k\frac{t}{\hbar }\left( 1-\frac{1}{\sqrt{t}}
\right) +\alpha \sqrt{t}}\exp \left( \displaystyle\frac{-\left(
y-x_{0}\right)^2} {2a^2}\right)\, dy\sim \\
&& \\
&&\displaystyle\frac1{\displaystyle\frac{1}{2}k \displaystyle \frac{t}{\hbar 
} \left( 1- \displaystyle\frac{1}{\sqrt{t}} \right) -\alpha \sqrt{t}}\exp
\left\{ \displaystyle \frac{-\left( \displaystyle\frac{1}{2}k \displaystyle%
\frac{t}{\hbar} \left( 1- \displaystyle \frac{1}{\sqrt{t}}\right) -\alpha 
\sqrt{t}-x_{0}\right) ^{2}}{2a^2}\right\} \rightarrow 0\quad \mbox{as}\quad
t\rightarrow \infty
\end{eqnarray*}
and the convergence rate is exponential in $t^{2}$. In the second interval
the error function can be expanded asymptotically for large $t$ as \cite
{Abramowitz} 
\begin{eqnarray*}
&&\frac{1}{4}i\frac{k}{\hbar ^{2}}me^{-\frac{1}{8}i\frac{k^{2}}{\hbar ^{3}}%
mt}\int_{-\frac{1}{2}k\frac{t}{\hbar }\left( 1-\frac{1}{\sqrt{t}}\right)
+\alpha \sqrt{t}}^{\infty } e^{-\frac{1}{2}i\frac{k}{\hbar ^{2}}my}\times \\
&& \\
&& \left[ 1-\mbox{erf}\left( -\frac{1}{4}i\frac{k}{\hbar ^{2}}m\sqrt{2}\sqrt{%
\left( i\hbar \frac{t}{m}\right) }+\frac{1}{2}y\frac{\sqrt{2}}{\sqrt{\left(
i\hbar \frac{t}{m}\right) }}\right) \right] \psi (y,0)\,dy \\
&& \\
&&\sim \frac{1}{4}i\frac{k}{\hbar ^{2}}me^{-\frac{1}{8}i\frac{k^{2}}{\hbar
^{3}}mt}\int_{-\frac{1}{2}k\frac{t}{\hbar }\left( 1-\frac{1}{\sqrt{t}}%
\right) +\alpha \sqrt{t}}^{\infty }\frac{e^{\frac{1}{2}i\frac{y^{2}}{\hbar t}%
m}}{2}\times \\
&& \\
&&\left\{ -\frac{k\sqrt{\hbar it}}{\sqrt{2\pi m}\left( ikt+y\right) }\left(
1-\sum_{n=1}^{\infty }\left( -1\right) ^{n}\frac{\prod_{j=1}^{n}\left(
2j-1\right) }{\left( 2\left( \frac{k}{2}\sqrt{\frac{\hbar it}{2m}}+y\sqrt{%
\frac{m}{2i\hbar t}}\right) ^{2}\right) ^{n}}\right) \right\} \psi (y,0)\,dy
\end{eqnarray*}
Expanding each term in the sum in powers of $y$ and using Watson's lemma, we
find that higher powers in $y$ produce lower powers of $t$ in the large $t$
asymptotic expansion of the integral. It follows that the leading order term
in the large $t$ expansion comes from the first term 1. Since $ikt+y\neq 0$,
we can evaluate the first term by extending the interval of integration over
the entire line with an error that decays exponentially fast in $t^{2}$. In
particular, setting $y=0$, we obtain the leading order contribution as 
\begin{eqnarray*}
&&\frac{1}{4}i\frac{k}{\hbar ^{2}}me^{-\frac{1}{8}i\frac{k^{2}}{\hbar ^{3}}%
mt}\int_{-\frac{1}{2}k\frac{t}{\hbar }\left( 1-\frac{1}{\sqrt{t}}\right)
+\alpha \sqrt{t}}^{\infty }\frac{e^{\frac{1}{2}i\frac{y^{2}}{\hbar t}m}}{2}%
\times \\
&& \\
&&\left\{ -\frac{k\sqrt{\hbar it}}{\sqrt{2\pi m}\left( ikt+y\right) }\left(
1-\sum_{n=1}^{\infty }\left( -1\right) ^{n}\frac{\prod_{j=1}^{n}\left(
2j-1\right) }{\left( 2\left( \frac{k}{2}\sqrt{\frac{\hbar it}{2m}}+y\sqrt{%
\frac{m}{2i\hbar t}}\right) ^{2}\right) ^{n}}\right) \right\} \psi (y,0)\,dy
\\
&& \\
&&\sim -\frac{1}{16}\frac{\sqrt[4]{\left( -1\right) }}{\hbar }\frac{\sqrt{2}%
}{\sqrt{\left( -ima^{2}+\hbar t\right) }}k\frac{\sqrt{m}}{\sqrt{\pi }}\exp
\left( \frac{1}{8}im\frac{4x_{0}^{2}\hbar ^{3}+ik^{2}tma^{2}-k^{2}t^{2}\hbar 
}{\left( -ima^{2}+\hbar t\right) \hbar ^{3}}\right)
\end{eqnarray*}
It follows that 
\begin{eqnarray*}
\phi \left( t\right) &\sim &\left( 1-i\right) \frac{\sqrt{m}}{2\sqrt{\left(
-ima^{2}+\hbar t\right) }\sqrt{\pi }}\exp \left( \frac{1}{2}ix_{0}^{2}\frac{m%
}{-ima^{2}+\hbar t}\right) - \\
&& \\
&&\frac{1}{16}\frac{\sqrt[4]{\left( -1\right) }}{\hbar }\frac{\sqrt{2}}{%
\sqrt{\left( -ima^{2}+\hbar t\right) }}k\frac{\sqrt{m}}{\sqrt{\pi }}\exp
\left( \frac{1}{8}im\frac{4x_{0}^{2}\hbar ^{3}+ik^{2}tma^{2}-k^{2}t^{2}\hbar 
}{\left( -ima^{2}+\hbar t\right) \hbar ^{3}}\right)
\end{eqnarray*}
Since the first term is not oscillatory for large $t$ and the second term is
oscillatory, we find that 
\[
\left| \phi \left( t\right) \right| ^{2}=O\left(t^{-1}\right)\quad%
\mbox{for large
$t$}. 
\]
It follows that the large $t$ asymptotics of $S(t)$ is 
\[
S(t)\sim 1-O\left(\ln t\right). 
\]
Higher order terms in the asymptotic expansion of $\phi \left( t\right) $
contribute integrable terms to $S(t)$. Thus $S(t)$ vanishes in finite time.

This means that the wave function vanishes in finite time everywhere outside 
$x=0$. That is, it develops a discontinuity in finite time. From that time
on, the expression eq.(\ref{psixt}), with $\psi(0,t)=\phi(t)$ given by eq.(%
\ref{phiti}), is no longer a solution of Schr\"odinger's equation. After
this time $\psi(x,t)$ vanishes identically outside $x=0$.

We conclude that the potential $-ik\delta (x)$ leads to total absorption of
the wave function in finite time.\newline

\noindent {\bf Acknowledgment:} The integrals were evaluated with Maple.%
\newline

\end{document}